\documentclass[doublecol]{epl2}

\usepackage{graphicx}
\usepackage{amssymb}
\usepackage{amsmath}
\usepackage[normalem]{ulem}

\usepackage{epstopdf}
\usepackage{color}

\title{A mean-field model of memristive circuit interaction}
\author{F. Caravelli\inst{1} \and P. Barucca\inst{2,3}}
\institute{$\ ^1$Theoretical Division and Center for Nonlinear Studies \\
Los Alamos National Laboratory, Los Alamos, New Mexico 87545, USA,   \email{caravelli@lanl.gov} \\
$\ ^2$Department of Computer Science, University College London, Gower Street, London WC1E 6BT, UK\and\\
$\ ^3$London Institute for Mathematical Sciences, 35a South Street, London W1K 2XF, UK}
\abstract{
We construct an exactly solvable circuit of interacting memristors and study its dynamics and fixed points. This simple circuit model interpolates between decoupled circuits of isolated memristors, and memristors in series, for which exact fixed points can be obtained. We introduce a Lyapunov functional that is found to be minimized along the non-equilibrium dynamics and which resembles a long-range Ising Hamiltonian with non-linear self-interactions. We use the Lyapunov functional as an Hamiltonian to calculate, in the mean field theory approximation, the average asymptotic behavior of the circuit given a random initialization, yielding exact predictions for the case of decay to the lower resistance state, and reasonable predictions for the case of a decay to the higher resistance state.}

\begin{document}
\maketitle

\section{Introduction}
Neuromorphic circuits are a promising technology to implement at the hardware level the computational power of analog computation inspired by the mammal brain. The type of computation performed by memristors requires a general theoretical understanding of the dynamics, in particular to allow controllability and interpretability of the results.
Memristors are becoming the most promising technology for the analogue implementation of artificial intelligence, and their dynamics is known to display memory effects \cite{diventra13a,pershin11d,festchua,chialvo}, being these very sensible to initial conditions \cite{indiveri,traversa13a,Stieg12}. Memristive circuits are also a new direction of study \cite{Caravelli2015,Caravelli2016} from a statistical physics standpoint, as these show critical behavior \cite{Caravelli2016rl,Sheldon} and can be connected to the solution of optimization problems \cite{traversa14b,Traversa2014,Caravelli2016ml}.
In its simplest description, a memristor is a 2-ports device behaving as a resistance which changes its value as a function of the flowing current. In this paper we restrict to ideal memristors with zero-crossing in the Voltage-Current diagram 
\cite{chua76a,stru08,stru12}, though more recently ReRAM (Resistive RAM) devices have further generalized this type of behavior \cite{Valov}. \\
The analysis presented in this paper could constitute a baseline for modelling and analyzing more complex interacting memristive circuits. We introduce a simple circuit whose asymptotic dynamics we show to be governed by a Lyapunov functional. As we will argue, such a functional can be casted into a spin-like model with long range interactions but with non-linear self-energy. The model we introduce interpolates in fact between a set of non-interacting memristive circuits and a single mesh of memristors. In a recent paper \cite{Caravelli2017}, it has been shown however that the interaction strength between memristors for generic circuits is controlled by the Hamming distance on the dual graph of the circuit. That paper however does consider memristors on a mesh, but \textit{between} meshes. Specifically, we consider the case in which the Hamming distance between each pair of memristors is one, hence representing a fully-connected model in which mean field techniques can be used. The source of our motivation for studying mean-field models of memristors is due to the difficulty of analyzing and understanding the behavior of general memristive circuits.
It is the purpose of this paper to show that the model we introduce  can be regarded as the analogue of the system studied in the well-known Curie-Wei\ss model for spin-spin magnetic interactions. The key difference is that in our case the single interacting element is not the spin of a particle but the internal memory of a memristors, characterized by a different functional form for the pairwise interaction and, most importantly, by a non-Hamiltonian dynamics.\\
The present paper is structured as follows: in Section 1, we revise the standard model of a simple circuit with one memristor, in Section 2, we define the model of interacting memristors, in which many a memristor are coupled with a central mesh characterized by a given conductance regulating the strengh of the interaction. In Section 3, we analytically and numerically characterize the circuit both in the case of deterministic and random initializations. Finally, in Section 4 we discuss the results and their implications both on real implementations of memristive circuits and on their theoretical modelling.

\section{Mean field memristive circuit}
We begin by introducing the memristors under study. As first observed in 
\cite{AtomicSwitch1}, physical memristors slowly relax to a limiting resistance even when a voltage is not applied. This observation implies that there is a competition between a phenomenon of decay and one of reinforcing, which is one of the key mechanisms for the learning ability of biological systems. 
We consider the simplest dynamical equation which captures such behavior.
The time evolution of a simple Ag+ memristor (atomic switch)\cite{AtomicSwitch2}:
\begin{equation}
\frac{d}{dt} w(t)=\alpha w(t)-\frac{R_{on}}{ \gamma } I = \alpha w(t)-\frac{ R_{on}}{ \gamma } \frac{S}{R(w)}
\label{eq:noise}
\end{equation}
where $0\leq w(t)\leq1$ is the internal memory parameter of the memristor, $R(w)= R_{on} (1-w)+R_{off} w$ is the resistance and $I$ and $S$ are the current and applied voltage respectively. Using this parametrization, $R_{on}$ and $R_{off}$ are the limiting resistances for $w=0$ and $w=1$ respectively ($R_{off}>R_{on}>0$). The constant parameters $\alpha$ and $\gamma$ set the timescales for the relaxation and excitation of the memristor respectively\footnote{In particular, meanwhile $\alpha$ has the dimension of an inverse time, $\gamma$ has the dimension of time and voltage.}. 
The fixed points $w^*$ can be obtained by setting $\frac{d}{dt} w=0$, from which we find the equation
\begin{equation}
 \frac{R(w^*)}{R_{on}} w^*= \left(  (1-w^*)+\frac{R_{off}}{R_{on}} w^* \right)w^*= \frac{S}{\alpha \gamma}.
\end{equation}
We immediately observe that this equation is quadratic in $w^*$ and that none, one, or two solutions can be obtained depending on the values of the parameters. Physical memristors relax to the state of highest resistance $R_{off}$ at zero voltage, i.e. $\alpha>0$. 

\begin{figure}
\centering
\includegraphics[scale=0.34]{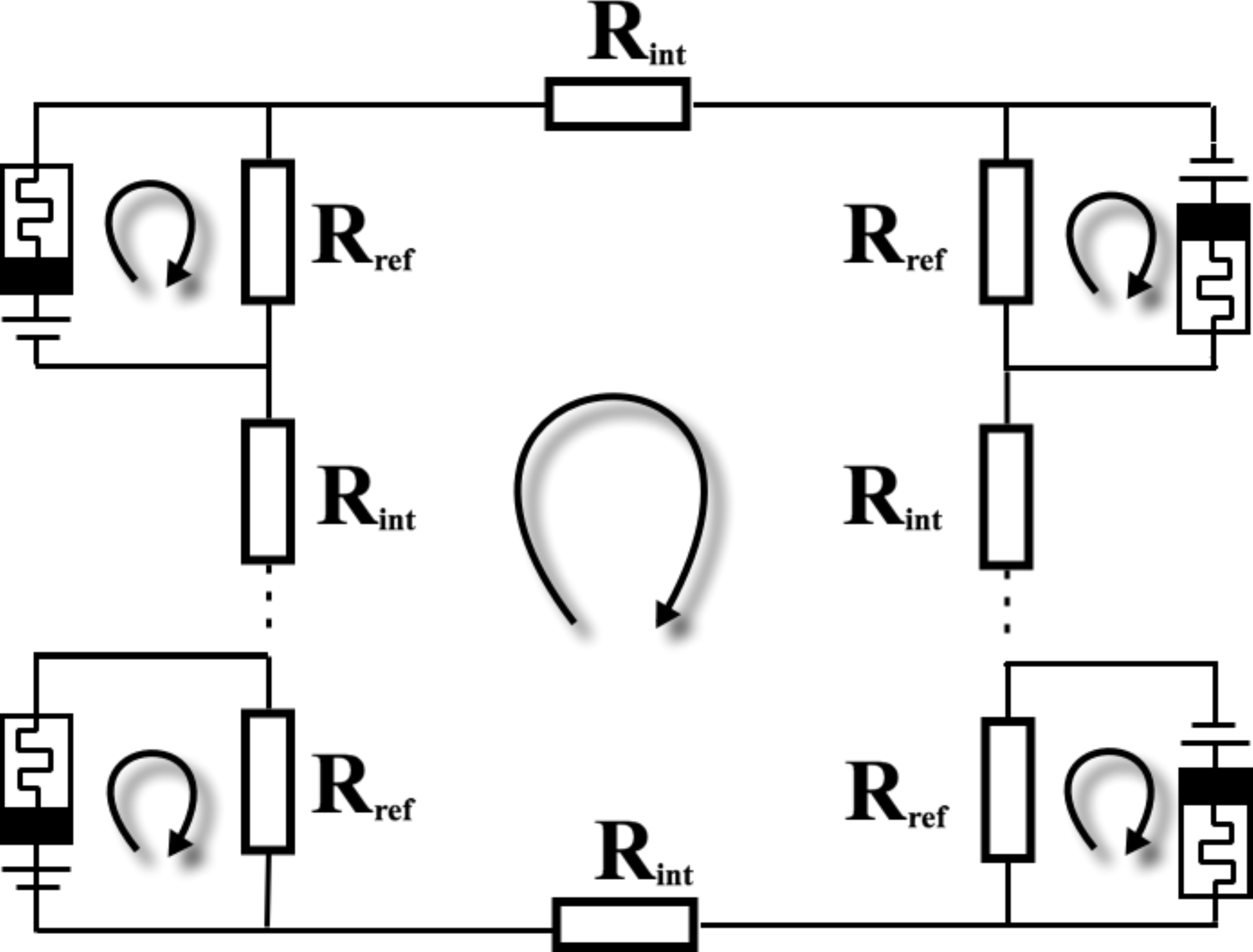}
\caption{More general case with $n$ memristors arranged on a mesh. Since the circuit is planar, all the meshs can be chosen with the same orientation.}
\label{fig:nmemr}
\end{figure}

The analysis above shows the main characteristics of memristive circuit dynamics: multiple fixed points depending on the value of the external control. With this feature in mind, we now generalize the dynamical model to the case of a circuit composed of $N$ memristors in series to a voltage source, as in Fig. \ref{fig:nmemr}: this is a simple modification of the one studied in \cite{carbajal} for machine learning purposes.  It was proven in \cite{Caravelli2017} that the interaction strength between memristors decays exponentially with the Hamming distance on the graph. Thus, in order to obtain long range interaction it is necessary to minimize the Hamming distance. The easiest way to obtain this is by arranging each memristor on a single mesh. However, since the memristor internal dynamics depends on the current, when these are arranged on a single mesh the dynamics of the circuit is trivial. The simplest non-trivial dynamics with long range interaction is obtained by inserting a current-divider on a central mesh, as in Fig. \ref{fig:nmemr}. Hence, we introduce control resistances in order to have long range interactions between memristors which are tunable. We consider $n$ memristors.
 For the derivation of the equations for the circuit of Fig. \ref{fig:nmemr}, we first apply the mesh current method \cite{Nilsson}, in which we assign to each memristive mesh a current $i_k$, $k=1,\cdots,n$, and we define the central mesh current  $i_0$. By construction, in the central mesh there are no memristors but only resistances. Once we have assigned an orientation to the mesh current, the current on each resistance is clear: on the $R_{ref}$ in parallel to the $k$th memristor, we will have a current $i_0-i_k$, meanwhile in any $R_{int}$ resistance will flow a current $i_0$. \\
On each $k$th memristor the current is given simply by $i_k$. 
The final step is to write the Kirchhoff voltage conservation law for each mesh. Since we have $n+1$ currents and $n+1$ meshes, we have a complete set of equations given by:
\begin{eqnarray}
\text{\textbf{Central mesh equation:}  }& &\nonumber \\
n R_{int} i_0(t)+ \sum_{k=1}^n R_{ref}(i_0(t)-i_k(t))&=&0,\\
\text{\textbf{Memristive meshes:} }& &\nonumber \\
R\left(w_k(t)\right) i_k+R_{ref}(i_k(t)-i_0(t))&=& S_k(t), \\
\text{\textbf{Memory evolution:} }& &\nonumber\\
\alpha w_k(t) -\frac{R_{on}}{\gamma} i_k(t)&=&\frac{d}{dt} w_k,
\label{eq:dynamics}
\end{eqnarray}
where $S_k(t)$ are voltage sources.
The system of equations \eqref{eq:dynamics} completely determines the circuit dynamics and yields the following memristors's dynamics, and can be derived via the Sherman-Morrison relation\cite{smf}:
\begin{equation}
\frac{dw_k}{dt} = \alpha w_k - \sum_{j=1}^n \frac{R_{on}}{\gamma}(\mathcal{I} - \frac{1}{n}M\mathcal{J} )_{kj}^{-1}\frac{S_j}{R_{ref}+R(w_j)},\label{eq:exp_dyn}
\end{equation}
where $M_{ij} = \delta_{ij}\frac{R_{ref}}{R_{ref}+R(w_i)}\frac{R_{ref}}{R_{ref}+R_{int}}$ and $\mathcal{J}$ is an all-ones matrix (see the Appendix of \cite{paperarxiv}) and $\mathcal{I}$ the identity matrix. 
In addition to the Kirchhoff laws above, we need to include the memory dynamics:
\begin{equation}
\frac{d}{dt} w_i = \alpha w_i -  f_i(\vec w), 
\end{equation}
 where, more specifically, $f_i(\vec w)=\sum_{j}g_i(w_i,w_j)$, which reflects the fact that the dynamics is non-linear, pairwise, and all-to-all. Each memristor interacts with every other, through an interaction term $g_i(w_i,w_j) = \frac{R_{on}}{\gamma}(\mathcal{I} - \frac{1}{n}M\mathcal{J} )_{ij}^{-1}\frac{S_j}{R_{ref}+R(w_j)}$ depending both on its own state $w_i$ at time $t$ and on the state of memristor $j$.
Via a direct computation of the partial derivatives, it is easy to see that the dynamics does not derive from a potential, as $\partial_{w_i} f_j(\vec w)\neq \partial_{w_j} f_i(\vec w)$. This means that we cannot strictly interpret the behavior of memristive circuits as an Hamiltonian dynamics. 
 In general, the statistical description of such non-Hamiltonian systems cannot restrict itself to the calculation of the equilibrium distribution.
In the following subsection, we identify a general approximation to map initial states into asymptotic states, depending on the stability of the fixed points of equations \eqref{eq:exp_dyn}. 
\section{Asymptotic dynamics}
In this Section we identify the general fixed point equations of the circuit's dynamics. 
The fixed points of this circuit can be obtained from eqns. (\ref{eq:dynamics}) if we set $\frac{d}{dt} w_k=0$. These equations generalize the fixed-point equation obtained for the single memristor. We derive in fact a direct relationship between equilibrium currents and the internal memory $w_k$:
\begin{eqnarray}
i_k&=&\frac{\alpha \gamma}{R_{on}} w_k  \\
i_0&=&\frac{\alpha \gamma }{R_{on}} \frac{R_{ref}}{ R_{int}+R_{ref}}   \frac{1}{n}\sum_{k=1}^n w_k\equiv \frac{\alpha \gamma }{R_{on}} \frac{R_{ref}}{ R_{int}+R_{ref}} \langle w \rangle \nonumber 
\end{eqnarray}
and then find a fixed point equation for the internal memory parameters, which are the solution of the following fixed point equations:
\begin{eqnarray}
\frac{S_k}{\alpha \gamma}&=&\frac{R(w_k)+R_{ref}}{R_{on}} w_k- \frac{R_{ref}^2}{ R_{on} (R_{int}+R_{ref})} \langle w \rangle. \nonumber
\end{eqnarray}
Considering that for an ideal memristor $R(w_k)=R_{off} w_k+(1-w_k) R_{on}=R_{on}+(R_{off}-R_{on})w_k$, we can rewrite this equation in terms of adimensional quantities only:
\begin{eqnarray}
\frac{S_k}{\alpha \gamma} &=&\frac{R_{off}-R_{on}}{R_{on}} w_k^2 \nonumber \\
 &+&\frac{R_{on}+R_{ref}}{R_{on}} w_k-\frac{R_{ref}^2}{ R_{on} (R_{int}+R_{ref})} \langle w \rangle \nonumber \\
 &=& \xi w_k^2 + \chi w_k -\rho \langle w \rangle,
 \label{eq:fixedpoint}
\end{eqnarray}
where we defined $\xi=\frac{R_{off}-R_{on}}{R_{on}}$, $\chi=\frac{R_{on}+R_{ref}}{R_{on}}$ and $\rho=\frac{R_{ref}^2}{ R_{on} (R_{int}+R_{ref})}$. 
We note that the mean internal memory $\langle w \rangle$ acts as an effective voltage source for the circuit.
We see that already for this rather simple circuit, all-to-all interactions can affect the position of the fixed points, as these are modified by a mean field term $ \langle w \rangle$ directly proportional to $\rho$.

\subsection{Asymptotic distributions for the homogeneous case}

\begin{figure}
\includegraphics[scale=0.23]{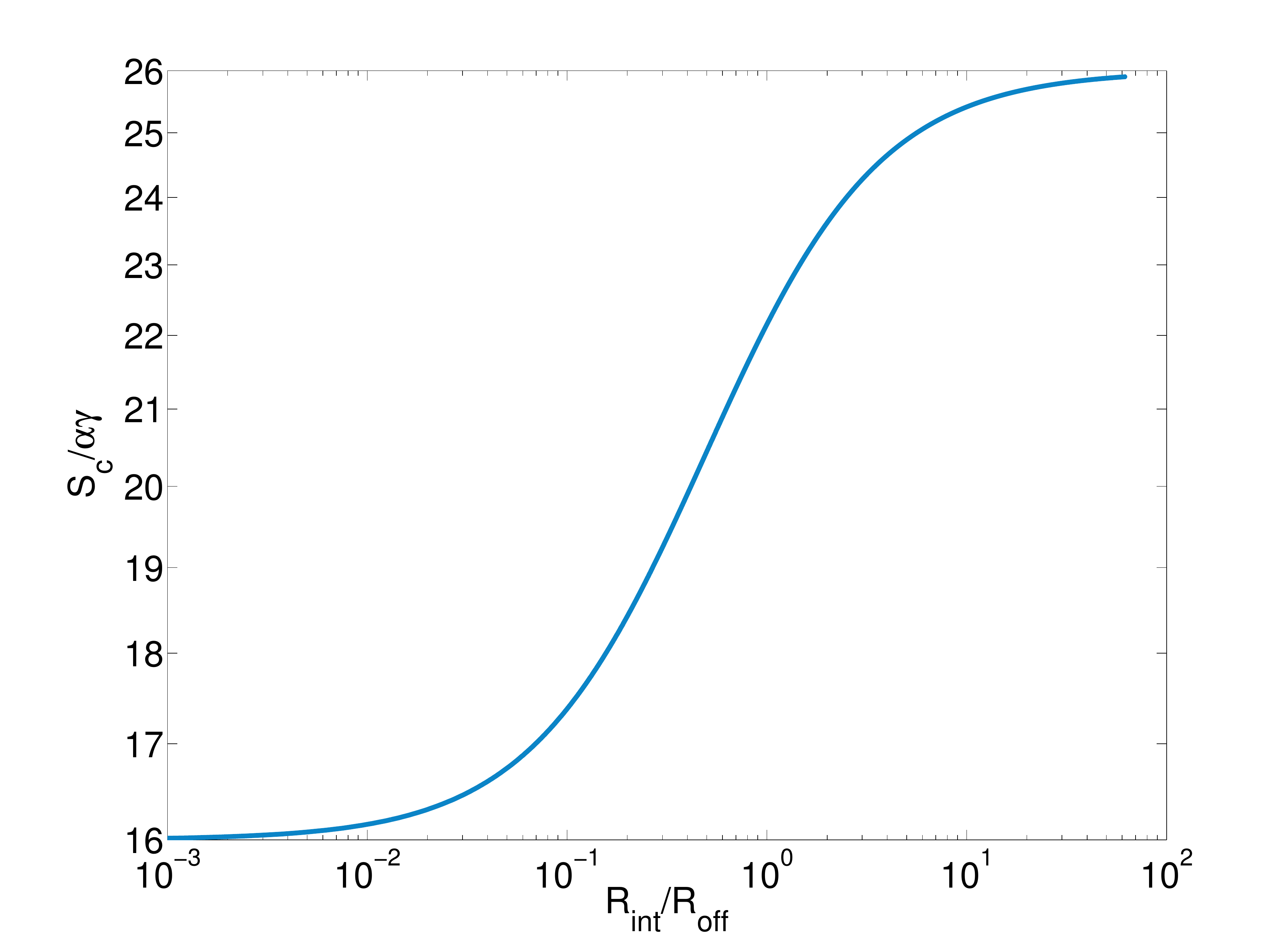}
\caption{Critical line of eqn. (\ref{eq:firstorder}) of the non-equilibrium dynamics starting from switched off memristors, $w_i(0)=1$. Parameters of the model: $R_{ref} = 1000$, $R_{on} = 100$, $R_{off} = 1600$, and $\alpha = \gamma = 1$. }
\label{fig:dyn0}
\end{figure}

Starting from the general dynamics and fixed-point equations, we consider the case of homogenous applied voltage across all memristors, $S_k=S$, and consider two different initial conditions. The non-equilibrium dynamics of the system will determine two radically different asymptotic states for these two cases.  In the first case, all memristors start from a switched-off condition, $w_k(0)=1$ for all $k$. Such uniform condition combined with the homogenous applied voltage results in an exact mean-field dynamics. This implies that all memristors behave alike. This type of dynamics allows us to exactly compute  the asymptotic state varying the applied voltage $S$. The critical value for the voltage $S$ can be worked out from eq. \eqref{eq:fixedpoint} and reads:
\begin{eqnarray}
S_c =\alpha \gamma \left( \frac{R_{off}+R_{ref}}{R_{on}}  -\frac{R_{ref}^2}{ R_{on} (R_{int}+R_{ref})}\right).
 \label{eq:firstorder}
\end{eqnarray}
Fig. \ref{fig:dyn0} shows the critical line as a function of the parameter $R_{int}/R_{off}$. Above that line, memristors are activated and all converge to the state of lower resistance $w=0$, corresponding to $R(w)=R_{on}$, while below the same line memristors will remain on the state of higher resistance $w=1$, corresponding to $R(w)=R_{off}$.  As we will see, this simple analysis shows the existence of a phase transition, that we will study using mean-field techniques.

\subsection{Non-interacting case} 
\begin{figure}
\includegraphics[scale=0.32]{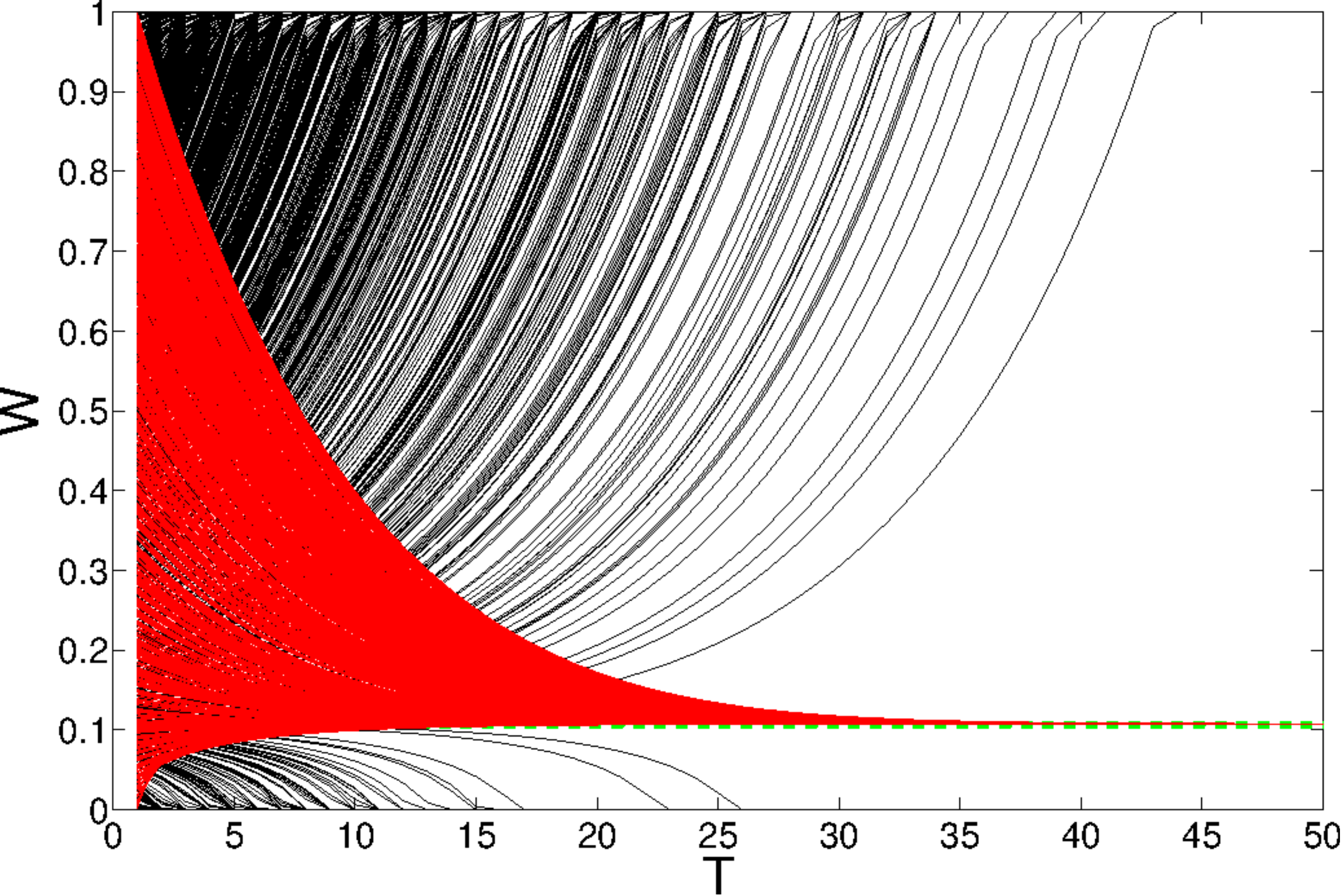}\\
\includegraphics[scale=0.32]{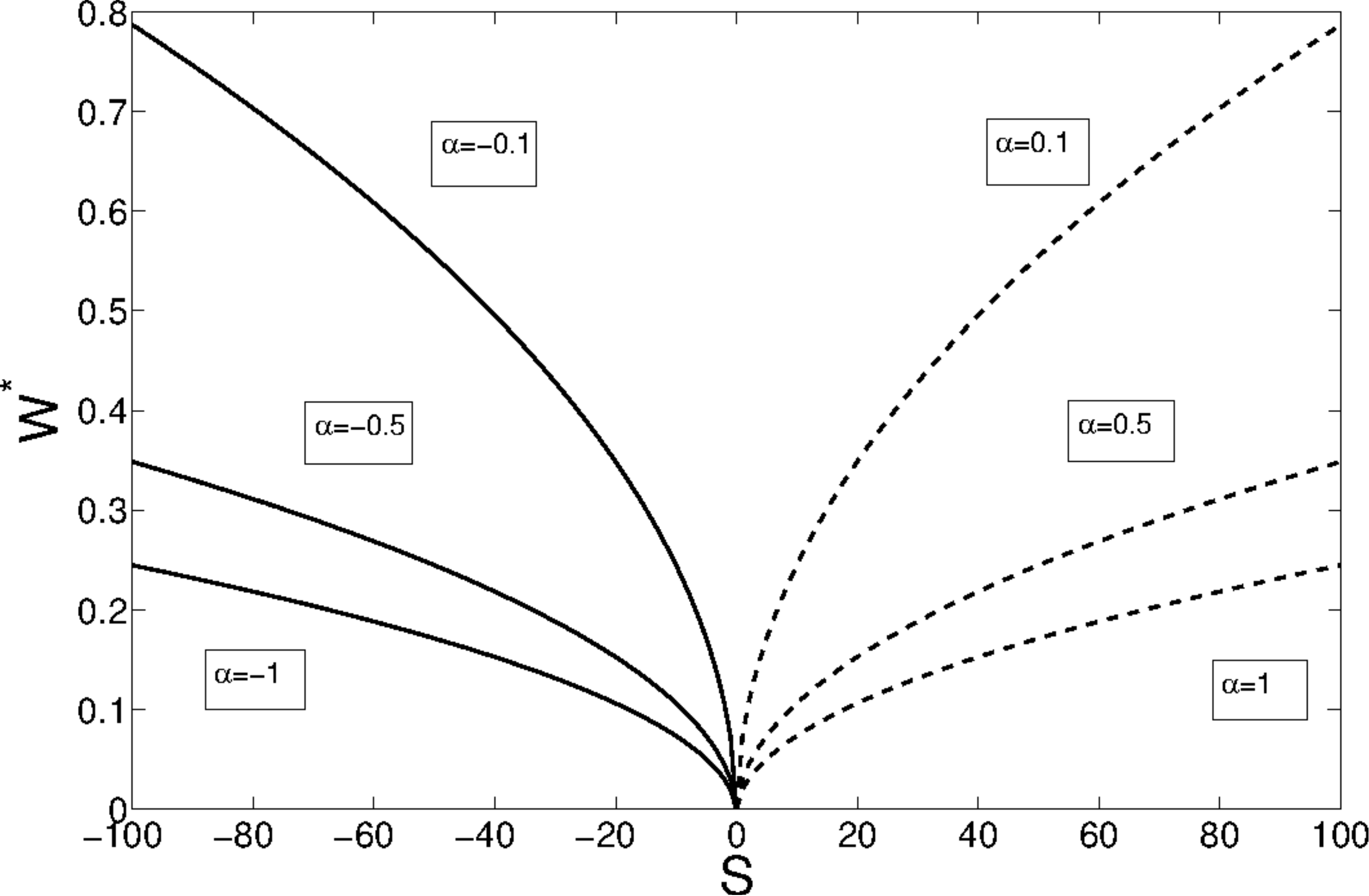}
\caption{\textit{Top:} Dynamics for the interacting case in the case $R_{ref}=100=R_{int}=100$, $R_{on}=100$, $R_{off}=16000$, $\gamma=10$. The case of $S=20$ and $\alpha=1$ is shown with black curves, meanwhwile $S=-20$, $\alpha=-1$ in red. These equations have been obtained solving numerically eqns. (\ref{eq:dynamics}) with an integration step $dt=0.1$ and $n=1000$ memristors. The blue dashed line is the threshold calculated from eqn. (\ref{eq:eom}). We note that the same fixed point can describe an attractive or a repelling fixed point, depending on the signs of $S$ and $\alpha$.\ \ 
\textit{Bottom:} Fixed points as functions of $\alpha,S>0$ are dashed curves (unstable) and $\alpha,S<0$  are continuous curves (stable).}
\label{fig:dyn}
\end{figure}

The case of a random initial condition is radically different as a homogeneous dynamics is not established by the system. In this case, we need to introduce a mean-field approximation of the heterogenouos dynamics of the memristors. This approximation provides an estimate of the asymptotic state of activation.
Let us consider first the case in which the memristors are non-interacting, i.e. we  assume that $\rho=0$. Using this parametrization, the non-interacting estimate becomes:
\begin{eqnarray}
 \frac{S}{\alpha \gamma} &=& \xi  w ^2+ \chi  w
 \label{eq:eom}
\end{eqnarray}
where $\xi$ and $\chi$ are the adimensional parameters defined in eqn. (\ref{eq:fixedpoint}), $\frac{S}{\alpha\gamma}$ can be tuned using the external voltage sources, and $\chi$ depends on the interaction between the memristive meshs. We note that $\xi$ is typically positive as $R_{off}\gg R_{on}$,  and that $\chi$ cannot be negative for any positive values of the resistances. In the case $R_{ref}= 0$, implying $\rho = 0$ and $\chi = 1$. The solution of equation \eqref{eq:eom} is:
\begin{equation}
 {w^*}_{\pm}=-\frac{\chi}{2\xi}\pm\sqrt{\frac{\chi^2}{4 \xi^2}+ \frac{S}{\xi \alpha \gamma}}=-\frac{1}{2\xi}\left(1\pm\sqrt{1+4\frac{S \xi}{\alpha \gamma }}\right)
 \label{eq:wpm}
\end{equation}
We observe that necessarily one root of eqn. (\ref{eq:wpm}) falls below zero, and thus only one solution is feasible. 
In Fig. \ref{fig:dyn} (top) we plot the numerical solutions obtained for $\alpha<0$ and $\alpha>0$. 
One important fact that we need to stress is that the dynamics of the circuits greatly depends on the signs of $S$ and $\alpha$. The case in which $S_i$'s and $\alpha$ have identical signs is, as the fixed points will not fall in the interval $[0,1]$, and thus the asymptotic state for memristors is binary, either $\{0,\,1\}$.
We observe that meanwhile for $\alpha<0$ the asymptotic fixed point is stable, in the case $\alpha>0$ (which is the physical case) it is unstable. A simple calculation of the Jacobian confirms this fact.
 Fig. \ref{fig:dyn} (bottom) shows the position of the fixed points as a function of $\alpha$ and $S$ for both values of $\alpha$.

\subsection{General case: use of Lyapunov functional}
In the case $\alpha<0,S>0$, the asymptotic fixed point can be described as the minimum of a functional. In fact, eqn. (\ref{eq:fixedpoint}) can be obtained from $\partial_{w_i} H=0$, where $H$ is given by:
\begin{equation}
H(w_i)=\frac{\rho}{2n}\sum_i w_i^2 - \frac{\rho}{ n} \sum_{i,j} w_i w_j - \sum_i  \frac{S_i}{\alpha \gamma} w_i +\sum_i E(w_i)
\label{eq:hamiltonian}
\end{equation}
and where $E_i(w_i)=\frac{\xi}{3}  w_i^3 +\frac{\chi}{2} w_i^2$. 
As we will see, $H(w_i)$ will be used as an approximate Hamiltonian for the Curie-Wei$\ss$ mean field theory.
\begin{figure}
\includegraphics[scale=0.32]{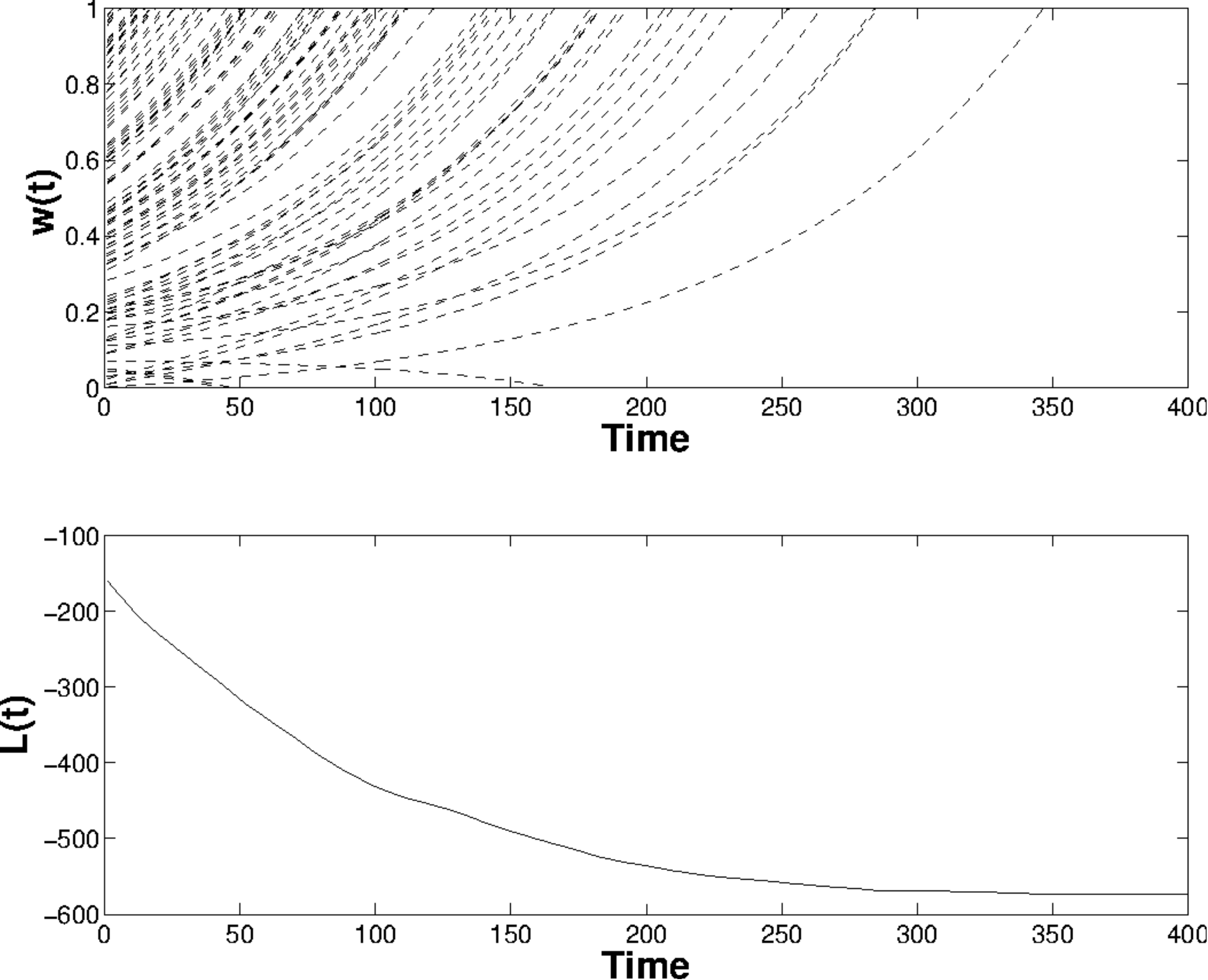}
\caption{The evolution over time of the Lyapunov function as the memristors converge to their asymptotic values.}
\label{fig:lyap}
\end{figure}
It is important to note that the functional in eqn. (\ref{eq:hamiltonian}) is a Lyapunov function, as we show in the Appendix of \cite{paperarxiv}, implying that $\frac{d}{dt} H(w_i)<0$ for $\alpha<0$. Thus, for $\alpha<0$ the asymptotic states $\langle w(t=\infty)\rangle$ are directly connected to fixed points of a functional which can serve as a Hamiltonian. For $\alpha>0$ the Lyapunov functional is given by $L=-H$ in eqn. (\ref{eq:hamiltonian}); in which the evolution of the Lyapunov functional is shown in Fig. \ref{fig:lyap}.
Although the Hamiltonian of eqn. (\ref{eq:hamiltonian}) is reminiscent of an Ising model with long-range interactions but with non-linear self-energy \cite{Campaetal,Campa}, it is worth to mention that here the parameters $w_i$ take values in $[0,1]$. To begin with, we anticipate that we will perform all the statistical mechanical calculations at a non-zero temperature, and then take the limit $T\rightarrow0$ at the end.

The first question we aim to answer is whether we can use the (unphysical) situation $S>0,\alpha<0$ to make any statement regarding the behavior of the system for $S<0,\alpha>0$. As mentioned before, $\alpha>0$ corresponds to a relaxation into an insulating phase, which is the physical case observed in $Ag+$ memristors \cite{AtomicSwitch1,AtomicSwitch2}. We can thus take advantage of a heuristic observation:  the fact that location of the unstable fixed determines which fixed point each memristor will reach. For instance,  in Fig. \ref{fig:dyn} (top), if $w_{in}>w^*$, since the time derivative is positive, we would expect $w(t=\infty)=1$; if on the other hand $w_{in}<w^*$, then we would expect $w(t=\infty)=0$, up to a set of measure zero, $w_{in}=w^*$. This also shows that there is a duality between the cases $S<0,\ \alpha>0$, and $S>0,\ \alpha<0$, which is also evident by the fact that since the fixed points depend only on the ratio $S/\alpha$, the position of the fixed point will be unaffected. We can then use the following rule of thumb which connects the probability on the asymptotic states for $\alpha<0$ to the ones of the initial states for $\alpha>0$:
\begin{eqnarray}
P(w(t=\infty)=1)&=& P(w_{in}>w^*) \nonumber \\
P(w(t=\infty)=0)&=& P(w_{in}<w^*) \nonumber \\
P(w_{in}<w^*)&=&1-P(w_{in}>w^*).
\label{eq:approxunstable}
\end{eqnarray}

With this in mind, we use the Lyapunov functional of eqn. (\ref{eq:hamiltonian}) as a Hamiltonian in a statistical mechanics setting. If we assume random initial conditions, we can try to predict  $\psi=\langle w(t=\infty) \rangle$ using an equilibrium approach in a canonical setting. Given the similarity of the functional to an Ising model, implies a Curie-Wei\ss\ approach for the average magnetization $\langle w \rangle$. Using standard mean field theory techniques and after some straightforward calculations (which are provided in the Appendix in \cite{paperarxiv}), we find the following mean field theory equation at zero temperature \cite{SchneiderPytte}:
\begin{equation}
\psi=  \text{arg sup}_{w\in[0,1]} \left( \left(\rho \psi+\frac{S}{\alpha \gamma}\right)w-E(w) \right).
\label{eq:mft}
\end{equation}
Eqn. (\ref{eq:mft}) can be exactly inverted as a function of $\psi=\langle w\rangle$. This gives the same result as the one we would obtain if we substituted $w_k\rightarrow \langle w\rangle$ in eqn. (\ref{eq:fixedpoint}). This is simply a correction to the physical parameter $\chi$, as in fact we obtain the same effective equation as the non-interacting approximation with the substitution $\chi\rightarrow \chi-\rho$, which is the correction due to the interaction between the memristors. Since the memristor memory is bounded between $0$ and $1$, we consider the function $\langle w\rangle=\text{max}\left(0,\text{max}\left(\psi(S),1\right)\right)$.

\begin{figure}
\includegraphics[scale=0.24]{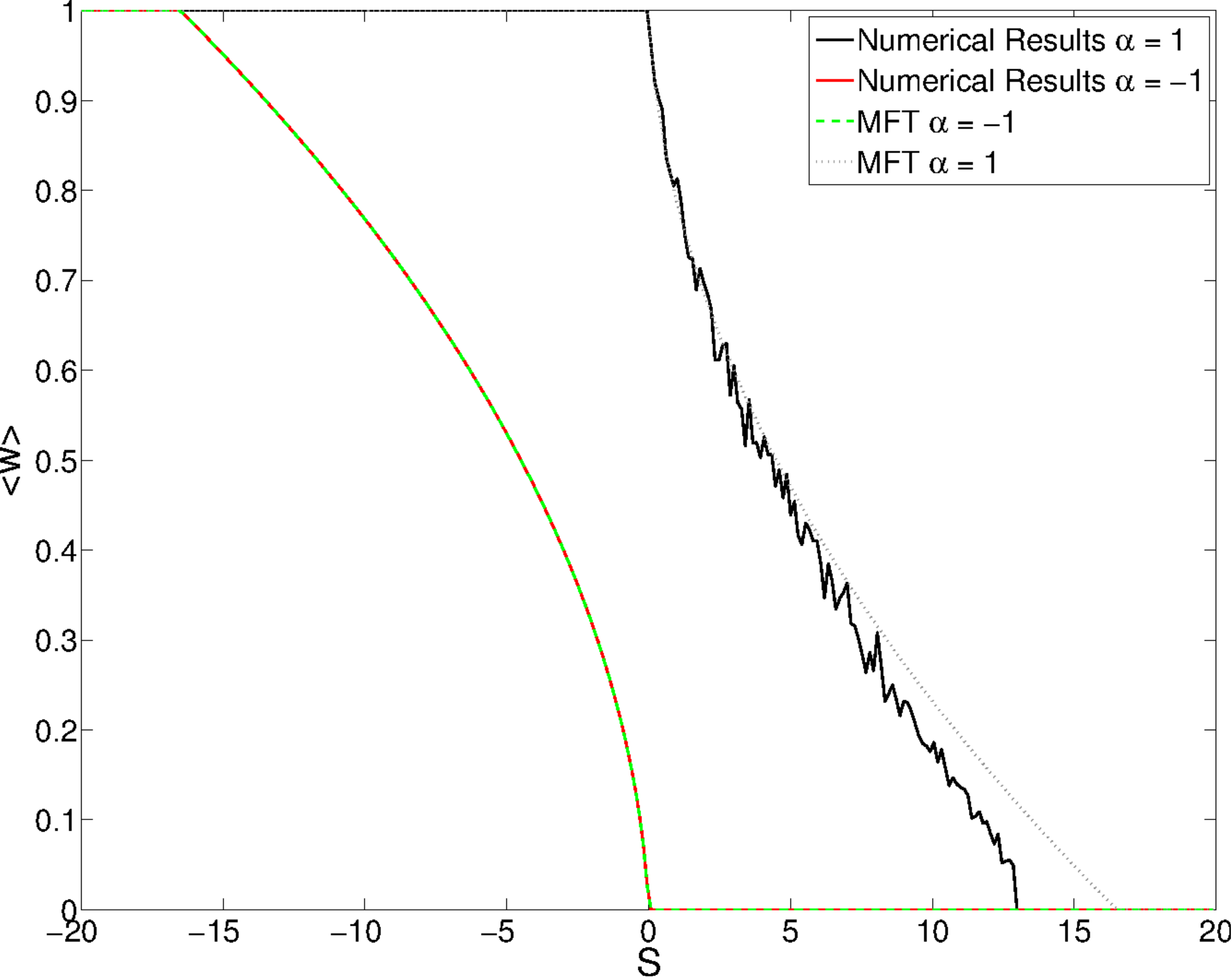}
\caption{Asymptotic fixed points as a function of $S$ for  $\alpha=\pm1$, $\gamma=1$, $R_{ref}=R_{int}=R_{on}=100$, $R_{off}=16000$ and $n=1000$ memristors. We compare the numerical results obtained by simulating the system and the theoretical estimate from the non-interacting assumption and the mean field theory, using the relation of eqn. (\ref{eq:approxunstable}). This figure has been obtained without averaging over the inital condition. For each point, the memristor memories were initiated randomly in $[0,1]$.}
\label{fig:solutions}
\end{figure}

In Fig. \ref{fig:solutions} we plot the numerical results on the mean field $\langle w(t=\infty)\rangle$ for $\alpha>0$ and compare these, as a function of $S$, to the non-interacting estimate obtained using eqn. (\ref{eq:approxunstable}). We observe that such approximation fails for larger values of $S$, but yet it provides nonetheless a good estimate for the asymptotic dynamics.
Few comments are in order. First we note that for $\alpha=-1$ the mean field theory calculation exactly reproduces the behavior of $\langle w \rangle$.
This approximation suggest a second order phase transition at $S=0$, as in fact one has $\partial_S \langle w \rangle$ given by
\begin{equation}
\partial_{S} \left(\text{max}\left(0,w^*(S) \right)\right)=
\begin{cases}
\frac{1}{\xi\alpha\gamma}\frac{1}{2 \sqrt{\frac{(\chi-\rho)^2}{4 \xi^2}+ \frac{S}{\xi \alpha \gamma}}} & S<0 \\
0 & S> 0
\end{cases}
\label{eq:phasetransition}
\end{equation}
which is not a differentiable function. 

For $\alpha>0$ we use the approximation of eqns. (\ref{eq:approxunstable}) to calculate the behavior of the system. The validity of this approximation is shown on the right hand side of Fig. \ref{fig:solutions}. We observe that the heuristic assumption of eqn. (\ref{eq:approxunstable}) for $\alpha<0$ is closer to the numerically obtained curve for $S\approx0$. For larger values of $S$ however, such approximation is less valid. We also observe stronger fluctuations around the mean field theory calculation, which we attribute to the effective instability of the fixed point. Nonetheless, such simple approximation provides a good estimate for the behavior of the internal memory also for larger values of the external applied voltage. The discrepancy is due to the fact that for larger values of $S$, because of the instability aforementioned, there can be trajectories which can invert their course. Since we observe that the real curve lies below the one obtained from the mean field theory, this implies that some memristors whose initial condition lies above the fixed point can invert and reach the asymptotic state $w=0$, rather than $w=1$.
We observe that eqn. (\ref{eq:approxunstable}) is valid for $S\approx0$, this implies that the discontinuity of eqn. (\ref{eq:phasetransition}) is valid also for $\alpha>0$. This seems evident from Fig. \ref{fig:solutions}, where the discontinuity in the first derivative of $\langle w\rangle$ between $\alpha<0$ and $\alpha>0$ is opposite in the sign of the derivative. Also, we note that possibly a divergence would occur if $\chi=\rho$ in eqn. (\ref{eq:phasetransition}), but this does not happen for any positive values of the resistances. 
From eqn. (\ref{eq:phasetransition}) we can also promptly infer the critical exponent of the transition point $S=0$ is $\frac{1}{2}$.

As a last comment, we note that we can derive the asymptotic susceptibility from the mean field exact solution in the approximation that $S(t)$ is stepwise constant.
The result is given by:
\begin{equation}
\partial_Sw(t\gg 1)\approx \frac{R_{on}}{2 \gamma \sqrt{\alpha (R_{off}-R_{on})}} e^{-\alpha t},
\end{equation}
and the derivation provided in the Appendix of \cite{paperarxiv}.


\section{Conclusions}
Memristors are becoming of interest for their use in on-chip implementations of neural networks \cite{neurm1,neurm2}.
Yet, only a few works focus on their collective dynamics. The results presented in this paper connect the Hamiltonian of interacting spin systems with the Lyapunov function of memristive circuits. This implies a connection between the equilibrium states of a statistical system of spins and the asymptotic states of memristors in a circuit. This statistical mechanics interpretation of memristors, is also analogous to the case of neural networks with temporal delays \cite{Herz}, and provides a solid basis for further generalizations.  It has been insofar hard to obtain an analytical control of the dynamics of memristive circuits, in this paper we introduced a mean field theory via a mapping between the equilibrium states into a suitable minima of Lyapunov function. Thus, we have shown, both analytically and numerically, that a first order phase transition occurs for positive $\alpha$, when memristors are initialized to the high-resistance state, and that a second order phase transition occurs when the initial conditions are chosen uniformly at random between the high and low resistance states. Our results on the mapping between non-equilibrium dynamics and equilibrium states could be extended to the more general case of purely memristive circuits. This work gives predictions on the convergence to boundary values of DC-controlled memristors, something which has already being introduced in \cite{Lyapunov} In future works we will consider the case of noisy memristive dynamics, heterogeneous applied voltages and a generic network topology \cite{Caravelli2016}, where complex non-equilibrium glass-like behaviour is expected \cite{Caravelli2017sg}.

\ \\\textbf{Acknowledgements.}
 The work of FC was carried out under the
auspices of the NNSA of the U.S. DoE at LANL under Contract No. DE-AC52-06NA25396. PB acknowledges support from FET Project DOLFINS nr. 640772  and FET IP Project MULTIPLEX nr. 317532;
PB and FC would like to thank the London Institute for Mathematical Sciences where part of this work was done.


\newpage
{\Large Supplementary Material\\}

\section{Single mesh toy model} \label{sec:inverse}
\subsection{Simple derivation}
From the mesh circuit equations we have: 
\begin{equation}
i_k = \frac{S_k}{R_{ref}+R(w_k)} +  \frac{R_{ref}}{R_{ref}+R(w_k)}i_0
\end{equation}
And, 
\begin{equation}
i_0 =  \frac{R_{ref}}{R_{ref}+R_{int}}\frac{1}{n}\sum_{k}^n i_k
\end{equation}
So that we have: 
\begin{equation}
i_k = \frac{S_k}{R_{ref}+R(w_k)} +  \frac{R_{ref}}{R_{ref}+R(w_k)}\frac{R_{ref}}{R_{ref}+R_{int}}\frac{1}{n}\sum_{j}^n i_j
\end{equation}
That leads to: 
\begin{equation}
i_k = \sum_{j=1}^n(\mathcal{I} - \frac{1}{n}M\mathcal{J} )^{-1}_{kj}\frac{S_j}{R_{ref}+R(w_j)}
\end{equation}
where $\mathcal{J}$ is the all-ones matrix and $M_{ij} = \delta_{ij}\frac{R_{ref}}{R_{ref}+R(w_i)}\frac{R_{ref}}{R_{ref}+R_{int}}$. 
And to the dynamics: 
\begin{equation}
\frac{dw_k}{dt} = \alpha w_k - \sum_{j=1}^n \frac{R_{on}}{\gamma}(\mathcal{I} - \frac{1}{n}M\mathcal{J} )_{kj}^{-1}\frac{S_j}{R_{ref}+R(w_j)}
\end{equation}
\subsection{Alternative derivation and formula}
An alternative derivation of the equation above can be obtained via a direct calculation of the inverse. The relation between currents and voltages can be  can be written as:
\begin{widetext}
\begin{equation}
\left(
\begin{array}{ccccc}
n(R_{int}+R_{ref}) & -R_{ref} & -R_{ref} & \cdots & -R_{ref} \\
-R_{ref} & R(w_1)+R_{ref} & 0 & \cdots & 0 \\
-R_{ref} & 0 & R(w_2)+R_{ref} & 0 & \vdots \\
\vdots & \vdots &  \ddots & \ddots & 0 \\
-R_{ref} & 0 & \cdots & 0 &  R(w_n)+R_{ref}
\end{array} \right)\left( \begin{array}{c} i_0 \\ i_1 \\ \vdots \\  \vdots \\ i_n \end{array}\right)=\left( \begin{array}{c} 0 \\ S_1 \\ \vdots \\  \vdots \\ S_n \end{array}\right)
\end{equation}
\end{widetext}
which means we need to invert a matrix of the form:
\begin{equation}
M=\left(
\begin{array}{ccccc}
a_0 & -b & -b & \cdots & -b \\
-b & a_1 & 0 & \cdots & 0 \\
-b & 0 & a_2& 0 & \vdots \\
\vdots & \vdots &  \ddots & \ddots & 0 \\
-b & 0 & \cdots & 0 &  a_n
\end{array} \right)
\label{eq:matriximp}
\end{equation}
which a special case of an arrowhead matrix.
The inverse of this matrix is rather complicated, but can be easily obtained by means of a cofactor formula:
$(A^{-1})_{ij}=\frac{1}{\text{det}(A)}C_{ji}$
where $C_{ij}=(-1)^{i+j}\text{det}(A^{\tilde i\tilde j})$ is the determinant of the matrix $A$ where the row $i$ and the column $j$ has been removed. We note that because of the properties of $A$, $C$ is a symmetric matrix.
Let us also note that the determinant of matrices of the form as in eqn. (\ref{eq:matriximp}), $D=\text{det}(M)=\prod_{k=0}^n a_k-b^2 \sum_{k=1}^n \prod_{j\neq k,j>0} a_j$.

Thus:
\begin{eqnarray}
(M^{-1})_{ij}=\frac{C_{ij}}{D}
\end{eqnarray}
We that the cofactor of the matrix in the case in which $i=j>1$ has the same form. We can thus already say that $C_{ij}$ for $i=j>1$ is of the form: $C_{ii}=\text{det}(M^{\tilde i \tilde i})=\prod_{k=0, k\neq i}^n a_k-b^2 \sum_{k=1,k\neq i}^n \prod_{j\neq k,j>0} a_j$.
The special cases $i=1,j>1$ and $j=1,i>1$, take the form $C_{1j}=C_{j1}=-b^2 \prod_{k\neq 1, k\neq j} a_k$. The case $C_{ij}$ 
with $i\neq j, i,j>1$ has to be calculated on its own. We note that for instance the matrix $C^{12}$ is of the form:
\begin{eqnarray}
C_{12}=C_{21}= -det\left(
\begin{array}{cccccc}
 a_0 & -b & -b & -b & -b & -b \\
 -b & 0 & 0 & 0 & 0 & 0 \\
 -b & 0 &a _3 & 0 & 0 & 0 \\
 -b & 0 & 0 &a _4 & 0 & 0 \\
 -b & 0 & 0 & 0 &a _5 & 0 \\
 -b & 0 & 0 & 0 & 0 & a_6 \\
\end{array}
\right)\\
\nonumber \\
\nonumber
\end{eqnarray}
and thus introduces a zero on the diagonal. For this reason, the determinant of these matrices are of the form $C_{ij}=-b^2 \prod_{k>0,k\neq i,k\neq j} a_k$.
We thus have:
\begin{widetext}
\begin{equation}
\left( \begin{array}{c} i_0 \\ i_1 \\ \vdots \\  \vdots \\ i_n \end{array}\right)=
\frac{1}{D} C \left( \begin{array}{c} 0 \\ S_1 \\ \vdots \\  \vdots \\ S_n \end{array}\right)\rightarrow \left( \begin{array}{c}  i_1 \\ \vdots \\  \vdots \\ i_n \end{array}\right)=
\frac{1}{D} 
\left(
\begin{array}{ccccc}
 q_1 & b^2\ c_{12} &\ b^2\ c_{13} & \cdots &\ b^2\ c_{1n} \\
 b^2\ c_{12} & q_2 & b^2\ c_{23}  &  & \vdots   \\
 b^2\ c_{13} & b^2\ c_{23} &\ddots &\ddots  &  \\
 \vdots &  & \ddots &  q_{n-1} &b^2\ c_{1 n-1}   \\
 b^2\ c_{1 n} & \cdots &   & b^2\ c_{1 n-1} & q_{n} \\
\end{array}
\right)
 \left( \begin{array}{c}  S_1 \\ \vdots \\  \vdots \\ S_n \end{array}\right)
\end{equation}
\end{widetext}
where we used the fact that because $S_0=0$ and we are also interested in $i_{k\geq 1}$, we can write the equation directly for the side loop. Also, we have implicitly defined:
\begin{eqnarray}
D&=&n (R_{int}+R_{ref})\prod_{k=1}^n (R(w_k)+R_{ref})\nonumber \\
&-&b^2 \sum_{k=1}^n \prod_{j\neq k,j>0} (R(w_k)+R_{ref})\\
c_{ij}&=&\prod_{k=1,k\neq i,j}^n (R(w_j)+R_{ref})\\
q_i&=&n (R_{int}+R_{ref})\prod_{k=1, k\neq i}^n (R(w_k)+R_{ref}) \nonumber \\
&-&b^2 \sum_{k=1,k\neq i}^n \prod_{j\neq k,i,j>0} (R(w_j)+R_{ref})\\
b&=&R_{ref}
\end{eqnarray}
We are in particular interested in the inverse of the submatrix which acts only on the memristor currents.
This can be easily obtained, and is given by $-\frac{R_{ref}^2}{n(R_{int}+R_{ref})} J+(R(\vec w)+R_{ref}) I$, where $J$ is the matrix made of ones and $R(\vec w)$ is the diagonal matrix with the resistances of each memristor. This implies the following dynamics:
\begin{eqnarray}
\frac{d}{dt} \vec w&=&\alpha \vec w  \nonumber \\
&-& \frac{R_{on}}{\gamma}\left(\text{diag}\left(R_{ref}+R(w)\right)- \frac{R_{ref}^2}{n(R_{int}+R_{ref})} J\right)^{-1}  \vec S.\nonumber\\
\label{eq:formaleq}
\end{eqnarray}
here $\text{diag}(\vec x)=\delta_{ij}x_j$, and which is the equation we find in the paper.

\subsection{Formal mean field dynamics solution and perturbative expansion}
In the mean field approximation, eqn. (\ref{eq:formaleq}) can be solved. If we use the Sherman-Morrison formula assuming that all memristors are equal, in such a case the equation becomes:
\begin{equation}
\frac{a s}{b w(t)+c}+w'(t)-\alpha  w(t)=0
\label{eq:meanfielddyn}
\end{equation}
with $a=\frac{R_{on}}{\gamma}$, $s=\langle \vec S\rangle=\frac{1}{n} \sum_{i=1}^n S_i$, $c=R_{ref}+R_{on}-\frac{R_{ref}^2}{R_{int}+R_{ref}}=R_{on}+\frac{R_{ref} R_{int}}{R_{ref} +R_{int}}$ and $b=R_{off}-R_{on}$, and where we are assuming that $S_i=\langle S_i \rangle$.

An analytic solution for such an equation can be found in terms of an inverse. Let us define:
\begin{eqnarray}
Q(t)&=&\frac{c\ \text{ArcTan}\left(\frac{\sqrt{\alpha } (2 (c_1+t)
   b+c)}{\sqrt{-4 a b s-\alpha  c^2}}\right)}{\sqrt{\alpha } \sqrt{-4 a b s-\alpha 
   c^2}} \nonumber \\
   &+&\frac{\log (a s-(c_1+t) \alpha  \left((c_1+t) b+c\right))}{2 \alpha} 
\end{eqnarray}
for an arbitrary integration constant $c_1$ due to time invariance symmetry. Then, the solution of eqn. (\ref{eq:meanfielddyn}) is given by the inverse function of $Q(t)$:
\begin{equation}
w(t)=Q^{-1}(t).
\label{eq:exactmfts}
\end{equation}
which is not analytical. In order to solve this equation, we use the a perturbative method in $\epsilon=c/b$, assuming that $R_{int}\gg R_{on}\gg1$. In fact, $\frac{1}{2}<\frac{R_{ref}R_{int}}{R_{ref}+R_{int}}<1$ for positive resistances. In this case, the differential equation becomes:
\begin{equation}
\frac{a s}{b w(t)}+w'(t)-\alpha  w(t)= \epsilon \frac{as}{b w(t)^2} +O(\epsilon^2)
\label{eq:meanfielddyn2}
\end{equation}
We thus search for perturbative solutions $w(t)=w_0(t)+\frac{c}{b}\ w_1(t)+\cdots$ up to the first order. We have the two differential equations up to the first perturbative order in $\epsilon$, which are:
\begin{eqnarray}
O(\epsilon^0)&:& \frac{a s}{b w_0(t)}+w_0'(t)-\alpha  w_0(t)= 0 \nonumber \\
O(\epsilon^1)&:& \frac{a s}{b w_0^2(t)}+w_1'(t)-\left(\alpha  +\frac{as}{b w_0^2(t)}\right)w_1(t)= 0 \nonumber \\
\end{eqnarray}
The zeroth perturbative order equation has solutions:
\begin{equation}
w_0(t)=\pm \frac{\sqrt{as+e^{2 \alpha  \left(b z_0+t\right)}}}{\sqrt{\alpha }
   \sqrt{b}}
\end{equation}
with $z_0$ associated with the initial condition, and of which we take the positive sign only. 
For $t>\frac{1}{2\alpha}$ we have an exponential function of the form:
\begin{equation}
w_0(t\gg 2\alpha)= \frac{e^{\alpha t}}{\sqrt{\alpha b}} +\frac{as e^{-2\alpha t}}{2\sqrt{\alpha b}} +O((as)^2)
\end{equation}
The first perturbative order is of the form:
\begin{equation}
x'(t)=h(t) x(t)+g(t)
\end{equation}
and has a general solution of the form:
\begin{equation}
x(t)=\left(e^{-\int^{ t } h(t^{\prime} ) dt^{\prime}}\right)
\left(z_1+\int^t g( t^\prime) e^{\int^{ t^\prime } h(t^{\prime\prime} ) dt^{\prime\prime} }dt^\prime\right) \end{equation}
where we identify $h(t)=\alpha+\frac{a s}{b w_0^2(t)}$ and $g(t)=\frac{a}{b w_0^2(t)}$. For large times $h(t\gg 2\alpha)\approx \alpha$ and $g(t\gg 2\alpha)\approx \frac{a}{b}\sqrt{\alpha b}^2 e^{-2\alpha t}=a\alpha e^{-2\alpha t}$.
Thus we have:
\begin{equation}
w_1(t\gg 2\alpha)\approx e^{-\alpha t}(z_1+ \frac{a}{b}e^{-\alpha t})
\end{equation}
We thus obtain the dynamic susceptibility for long times, which is given by:
\begin{eqnarray}
\partial_s w(t\gg2\alpha)&\approx&\frac{a}{2 \sqrt{\alpha } \sqrt{b} \sqrt{a s+e^{2 \alpha  \left(b 
   c_1+t\right)}}}\nonumber \\
   &\rightarrow& \frac{a}{2\sqrt{\alpha b}} e^{-\alpha t} \nonumber \\
   &=& \frac{R_{on}}{2 \gamma \sqrt{\alpha (R_{off}-R_{on})}} e^{-\alpha t}
\end{eqnarray}
which falls off exponentially in time.
\subsection{Fixed point structure}

Since the interaction matrix is always invertible, we can study the equivalent equation:
\begin{widetext}
\begin{equation}
\left(\text{diag}\left(R_{ref}+R(w)\right)- \frac{R_{ref}^2}{n(R_{int}+R_{ref})} J\right)\frac{d}{dt} \vec w=\alpha \left(\text{diag}\left(R_{ref}+R(w)\right)- \frac{R_{ref}^2}{n(R_{int}+R_{ref})} J\right) \vec w -\frac{R_{on}}{\gamma}   \vec S
\end{equation}
\end{widetext}
which we note we can rewrite the fixed point equation as:
\begin{equation}
0=\frac{1}{\alpha} \frac{d}{dt}\left( \chi w_i +\frac{\xi}{2} w_i^2 -\rho \langle w \rangle \right)= \chi w_i+\xi w_i^2- \rho \langle w \rangle -\frac{S_i}{\alpha \gamma}
\end{equation}
Fixed points then require that the following two equations are satisfied at the same time:
\begin{eqnarray}
 \chi w_i +\frac{\xi}{2} w_i^2 -\rho \langle w \rangle&=&c\\
 \chi w_i+\xi w_i^2- \rho \langle w \rangle -\frac{S_i}{\alpha \gamma}&=&0
\end{eqnarray}
for an arbitrary constant $c$ and with $w_i\in [0,1]$. 
\subsection{Lyapunov function}
Let us now consider the function
\begin{equation}
H(w_i)=\frac{\rho}{2n}\sum_i w_i^2- \frac{\rho}{n} \sum_{i, j} w_i w_j - \sum_i  \frac{S_i}{\alpha \gamma} w_i +\sum_i E(w_i)
\end{equation}
we have
\begin{eqnarray}
\frac{d}{dt} H&=&\sum_{i} (\partial_{w_i} H) \frac{d w_i}{dt}\nonumber \\
&=&\sum_{i,j} \left(-\rho \langle w \rangle-\frac{S_i}{\alpha \gamma}+\frac{\partial E}{\partial w_i}\right) \delta_{ij} \frac{d w_j}{dt}
\end{eqnarray}
and using the equations of motion we obtain
\begin{eqnarray}
\frac{d}{dt} H&=&\frac{1}{\alpha}\sum_{i,j}\frac{d w_i}{dt} \Big( (1+\frac{R_{ref}}{R_{on}}+\frac{R_{off}-R_{on}}{R_{on}} w_i)\delta_{ij}  \nonumber \\
&-&\frac{R_{ref}^2}{R_{on}(R_{ref}+R_{int})} \frac{1}{n} J_{ij}\Big) \frac{d w_j}{dt}
\end{eqnarray}
Let us define 
\begin{eqnarray}
Q_{ij}&=&\Big( (1+\frac{R_{ref}}{R_{on}}+\frac{R_{off}-R_{on}}{R_{on}} w_i)\delta_{ij} \nonumber \\
&-&\frac{R_{ref}^2}{R_{on}(R_{ref}+R_{int})} \frac{1}{n} J_{ij}\Big).
\end{eqnarray}
 We have:
\begin{equation}
\frac{d}{dt} H=\frac{1}{\alpha}\langle Q \frac{d}{dt} \vec w,\frac{d}{dt} \vec w\rangle=\frac{1}{\alpha}\langle \frac{d}{dt} \vec w,\frac{d}{dt} \vec w\rangle_{Q}=\frac{1}{\alpha}||\frac{d}{dt} \vec w||_{Q}^2
\end{equation}
This quantity is positive or negative depending on the sign of $\alpha$ and the eigenvalues of $Q$. If $Q$ is positive definite, then the sign of $\frac{d}{dt} H$ depends only on $\alpha$. We note that $Q$ is the sum of two Hermitean matrices. Thus the minimum eigenvalue of $Q$ satisfies the bound $\lambda_{min}(A+B)\geq \lambda_{min}(A)+\lambda_{min}(B)$ for $A$ and $B$ Hermitean. Thus:
\begin{eqnarray}
\lambda_{min}(Q)&\geq& \lambda_{min}\left((1+\frac{R_{ref}}{R_{on}}+\frac{R_{off}-R_{on}}{R_{on}} w_i)\delta_{ij}\right)\nonumber \\
&+&\lambda_{min}\left( -\frac{R_{ref}^2}{R_{on}(R_{ref}+R_{int})} \frac{1}{n} J_{ij}\right)
\end{eqnarray} 
Since $\frac{1}{n} J_{ij}$ has maximum eigenvalue $1$, we immediately observe that, since $R_{on}$ is positive by construction:
\begin{eqnarray}
\lambda_{min}(Q)&\geq1&+\frac{R_{ref}}{R_{on}}-\frac{R_{ref}^2}{R_{on}(R_{ref}+R_{int})} \nonumber \\
 &=&1+\frac{R_{ref}}{R_{on}}\left(1-\frac{R_{ref}}{R_{ref}+R_{int}}\right) \geq 1\  \nonumber \\
 & &\forall R_{int},R_{ref}\geq0
\end{eqnarray}
This implies that $Q$ is positive definite.
The function $H$ is thus a decreasing function of the dynamics when $\alpha<0$. Since the function is a weighted negative norm of the derivative of the memristor memories, then it is also zero at the fixed point. For $\alpha>0$, it is sufficient to define $-H$ as a Lyapunov function. In this case, the fixed points become $w=1$ and $w=0$.
\section{Mean field theory for random inizialization of homogeneous memristors}
We are interested in the low temperature regime of this model. Its partition function can be written as:
\begin{equation}
Z(\beta,n,S)=\text{Tr}_{w} e^{-\beta H(w)}
\end{equation}
with $H(w)$ from eqn. (\ref{eq:hamiltonian}),  $\beta=1/T$ and we have implicitly defined the trace:
$$\text{Tr}_{w}\left(\cdot\right)\equiv \prod_{i=1}^n \int_{0}^1 dw_i \left(\cdot\right).$$
We now use the Hubbard-Stratonovich identity, with $m=\frac{\sum_i \sigma_i}{n}$,
\begin{equation}
e^{bm^2}=\sqrt{\frac{b}{\pi}} \int_{-\infty}^\infty dx\ e^{-b x^2+2 mb x}
\end{equation}
with $b=\frac{n \rho \beta}{2}$. Let us define $\tilde E(w_i)=E(w_i)+\frac{\rho}{2n}\sum_i w_i^2$. We write:
\begin{eqnarray}
Z(\beta,n,S)&=&\text{Tr}_{w} e^{-\beta  \sum_{i=1}^n (\tilde E(w_i)-w_i \frac{S_i}{\alpha \gamma})} \sqrt{\frac{n \beta \rho}{2 \pi}}\cdot \nonumber \\
&\cdot& \int_{-\infty}^\infty d\psi\ e^{-\frac{n \beta \rho}{2} \psi^2+m n \beta \rho \psi} \nonumber \\
&=& \sqrt{\frac{n \beta\rho}{2 \pi}} \int_{-\infty}^\infty d\psi\ e^{-\frac{n \beta \rho}{2} \psi^2} Q(\beta,S,\psi)^n \nonumber \\
&=& \sqrt{\frac{n \beta \rho}{2 \pi}} \int_{-\infty}^\infty d\psi\ e^{-\frac{n \beta \rho}{2} \psi^2 +n \log\left( Q(\beta,S,\psi)\right) } 
\end{eqnarray}
where $Q(\beta,S,\psi)=\text{Tr}_{w} e^{\beta \left((\rho \psi  +\frac{S}{\alpha \gamma}) w -  \tilde E(w)\right)}$.If we take the limits $n\rightarrow \infty$ first, for which $\lim_{n\rightarrow\infty} \tilde E(w_i)=E(w_i)$, which gives
\begin{equation}
Z\approx e^{n\beta \tilde f(\beta)}
\end{equation}
with $f(\beta)=\text{arg min}_\psi \left(\frac{1}{2}\rho \psi^2-\frac{1}{\beta}\log\left(Q(\psi)\right)\right)$.

In turn, $f(\beta)$ is given by $\psi$ solution of
\begin{equation}
\rho \psi=\partial_\psi \left[\frac{1}{\beta} \log Q\left(\beta, w(\psi,S)\right)\right]=\frac{1}{\beta} \frac{\partial_\psi Q\left(\beta, w(\psi,S)\right)}{Q\left(\beta, w(\psi,S)\right)}
\end{equation}
Now we have
\begin{eqnarray}
\frac{1}{\beta} \frac{\partial_\psi Q\left(\beta, w(\psi,S)\right)}{Q\left(\beta, w(\psi,S)\right)}&=&\frac{\rho \beta}{\beta} \frac{\text{Tr}_{w}  w e^{\beta \left((\rho \psi  +\frac{S}{\alpha \gamma}) w -  E(w)\right)} }{\text{Tr}_{w} e^{\beta \left((\rho \psi  +\frac{S}{\alpha \gamma}) w -  E(w)\right)}}\nonumber \\
&=&\rho  \frac{\frac{1}{\beta}\text{Tr}_{w}  w e^{\beta \left((\rho \psi  +\frac{S}{\alpha \gamma}) w -  E(w)\right)} }{\frac{1}{\beta}\text{Tr}_{w} e^{\beta \left((\rho x  +\frac{S}{\alpha \gamma}) w -  E(w)\right)}}
\end{eqnarray}
which, in the limit $\beta\rightarrow \infty$ is given by the following mean field equation:
\begin{eqnarray}
\psi&=& \text{arg sup}_{w\in[0,1]} \left( \left(\rho \psi+\frac{S}{\alpha \gamma}\right)w-E(w) \right) \nonumber \\
&=&\sqrt{\frac{\chi ^2}{4 \xi ^2}+\frac{\frac{S}{\alpha  \gamma }+\rho  \psi}{\xi
   }}-\frac{\chi }{2 \xi }
\end{eqnarray}
which is the result presented in the paper.
The mean field susceptibility at equilibrium can be calculated from this equation. We have:
\begin{eqnarray}
\langle w^2\rangle&=&\frac{1}{\beta}\frac{\partial \psi}{\partial S}=\frac{1}{\alpha \gamma}  \frac{\frac{1}{\beta}\text{Tr}_{w}  w^2 e^{\beta \left((\rho \psi  +\frac{S}{\alpha \gamma}) w -  E(w)\right)} }{\frac{1}{\beta}\text{Tr}_{w} e^{\beta \left((\rho x  +\frac{S}{\alpha \gamma}) w -  E(w)\right)}} \nonumber \\
&\underbrace{\approx}_{\beta\rightarrow \infty}& \frac{1}{\alpha \gamma} \left(\text{arg sup}_{w\in[0,1]} \left( \left(\rho \psi+\frac{S}{\alpha \gamma}\right)w-E(w) \right) \right)^2\nonumber \\
&=&\frac{\psi^2}{\alpha \gamma},
\end{eqnarray}
which is the result shown in the main text.

\end{document}